# A Multi-Modal Miniature Surface Forces Apparatus (μSFA) for Interfacial Science Measurements

*Kai Kristiansen[‡,*,9], Stephen H. Donaldson Jr[†,9], Zachariah J. Berkson[‡,Ф], Jeffrey Scott[§], Rongxin Su[⊥], Xavier Banquy[¶], Dong Woog Lee[#], Hilton B. de Aguiar[†], Joshua D. McGraw[†,∣], George D. Degen[‡], and Jacob N. Israelachvili[‡].*

[‡]Department of Chemical Engineering, University of California Santa Barbara, Santa Barbara, CA93106, United States

[†]Département de Physique, Ecole Normale Supérieure/PSL,Research University, CNRS, 24 rue Lhomond, 75005 Paris, France

[§]SurForce LLC, Goleta, CA, 93117, United States

[⊥]State Key Laboratory of Chemical Engineering, Tianjin Key Laboratory of Membrane Science and Desalination Technology, School of Chemical Engineering and Technology, Tianjin University, Tianjin 300072, China

[¶]Faculty of Pharmacy, Université de Montréal, Succursale Centre Ville, Montréal Quebec H3C 3J7, Canada

[#]School of Energy and Chemical Engineering, Ulsan National Institute of Science and Technology, Ulsan 44919, Republic of Korea

[∣] Gulliver CNRS UMR 7083, PSL Research University, ESPCI Paris, 10 rue Vauquelin, 75005 Paris, France

[9]KK and SHD contributed equally to this work.

ABSTRACT

Advances in the research of intermolecular and surface interactions result from the development of new and improved measurement techniques and combinations of existing techniques. Here, we present a new miniature version of the Surface Force Apparatus – the µSFA – that has been designed for ease of use and multi-modal capabilities with retention of the capabilities of other SFA models including accurate measurement of surface separation distance and physical characterization of dynamic and static physical forces (i.e., normal, shear, and friction) and interactions (e.g., van der Waals, electrostatic, hydrophobic, steric, bio-specific). The small physical size of the µSFA, compared to previous SFA models, makes it portable and suitable for integration into commercially available optical and fluorescence light microscopes, as demonstrated here. The large optical path entry and exit ports make it ideal for concurrent force measurements and spectroscopy studies. Examples of the use of the µSFA in combination with surface plasmon resonance (SPR) and Raman spectroscopy measurements are presented. Due to the short working distance constraints associated with Raman spectroscopy, an interferometric technique was developed and applied for calculating the inter-surface separation distance based on Newton's rings. The introduction of the µSFA will mark a transition in SFA usage from primarily physical characterization to concurrent physical characterization with *in situ* chemical and biological characterization to study interfacial phenomena, including (but not limited to) molecular adsorption, fluid flow dynamics, determination of surface species and morphology, and (bio-)molecular binding kinetics.

INTRODUCTION

Interfaces are ubiquitous in nature and pursuits to control, understand, and modify interactions across interfaces have spanned millenia.[1] Advances in fundamental and applied nanoscience and surface science research[2-3] in recent years have profoundly impacted technology, medicine,[4] and our fundamental insight into the physical world.

To gain a better fundamental understanding of intermolecular and inter-particle interactions at surfaces, there have been numerous efforts to combine multiple measurement techniques into single instruments, including Atomic Force Microscopy with Electrochemistry,[5] tip-enhanced Raman spectroscopy,[6] Quartz Crystal Microbalance – Dissipation with Reflectometry,[7] Surface Plasmon Resonance with Electrochemistry,[8] Magnetic Resonance Imaging-Photoacoustic Imaging-Raman Imaging,[9] and Surface Forces Apparatus with X-ray spectroscopy[10] and electrochemistry[11]. Combined techniques allow concurrent *in situ* measurements– a requirement for sensitive systems in which small variations can have a huge impact on the physical and chemical properties. An example is the interaction between cytoplasmic cell membranes. The distance between cytoplasmic cell membranes in the human body is on average no more than 10 nm. At such short separation distances between two objects, the interaction energies (e.g., electrostatic attraction or repulsion, van der Waals and hydrophobic, and complementary (bio-)specific interactions) induce rearrangements of constituent species within each cell membrane that would otherwise not occur within isolated membranes. Only minor changes in the lipid/protein composition or the presence of very small quantities of additives such as polymers or nanoparticles can dramatically change membrane domain morphologies, with correspondingly dramatic influences on adhesion and other interactions between membranes.[12-14] Measuring and understanding the influences of such subtle effects often requires combined techniques applied under *in situ* conditions, providing motivation for the development of new combined instruments.

The Surface Forces Apparatus (SFA) has been a workhorse for interfacial science for the last fifty years.[15] The continuous development of the commercially available SFA, the SFA2000,[16] has allowed simultaneous force-measurements with (i) electrochemistry for precise control of surface potential and current flow,[11, 17] (ii) fluorescence microscopy to study simultaneous membrane morphological changes and intermembrane energies and adhesion,[18] and (iii) a 3D sensor-actuator for measuring anisotropic friction forces.[19] A specialized SFA system has been developed for simultaneous X-ray characterization.[10] Another SFA system has been developed for simultaneous capacitive detection to measure viscoelastic properties and electric field response of thin films, and properties at the solid-liquid interface (e.g., slippage, adsorbed liquid layers).[20] While the SFA2000 was designed to accept numerous attachments (e.g., friction, electrochemistry, fluorescence), it is too bulky to easily interface with typical laboratory microscope systems and thereby leverage many widespread microscopy techniques. Therefore, a new miniature Surface Forces Apparatus (μSFA, shown in Figure 1) has been designed that, like the SFA2000, enables accurate and unambiguous force-distance measurements and additionally enables simultaneous optical imaging and spectroscopic characterization of interacting surfaces in real time.

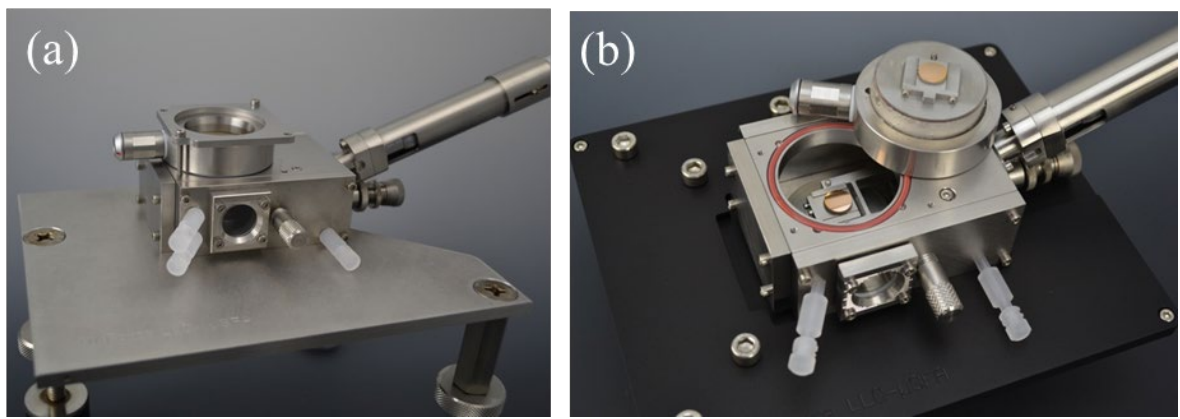

**Figure 1.** The miniature Surface Forces Apparatus (μSFA) on a (a) chamber maintenance stand and (b) custom microscope stage for an Olympus microscope. In (b) the piezo top mount is inverted to show the two mica surfaces back-coated with gold layer. The length of the front (and back) panel is 82 mm, and the chamber height is 48 mm at the piezoelectric upper surface attachment and 34 mm high everywhere else (excluding the motor drive assembly). The compact size of the μSFA allows for easy interfacing with common commercial optical light microscopes.

The μSFA can perform the same force measurements as previous SFA instruments and the large ports allow it to interface with diverse optical and spectroscopic techniques including fluorescence imaging, Raman spectroscopy, X-ray diffraction and/or spectroscopy, and Surface Plasmon Resonance (SPR) spectroscopy. The μSFA is ideal for studying complex surface interactions, processes, and morphologies in nano- to meso-scale systems (both biological and non-biological). The instrument enables simultaneous dynamic characterization of surface layer morphologies, processes, and interactions and is expected to yield fundamental insights into the relationship between changes in microscopic and submicroscopic structure and organization, molecular binding energies and kinetics, and inter-surface interactions.

Here, we describe the µSFA and demonstrate its capabilities for several model cases. Normal and frictional force measurements were conducted with simultaneous fluorescence microscopy, SPR spectroscopy, or Raman spectroscopy. The Raman-µSFA was especially challenging due to the short working distance constraints, which were mitigated by developing an interferometric technique for calculating the surface separation distance based on Newton's rings.

MATERIALS AND METHODS

**Chemicals and Materials.** Muscovite mica pieces of high optical grade (S&J Trading) were cleaved to 1-5 µm thin sheets before being placed onto a freshly cleaved mica backing sheet.[21] Silver was deposited on one side of the mica in a thermal evaporator, and the sheets were stored in desiccators at low pressures until used. Before use the mica sheets were peeled off the backing sheet and glued (thermosetting glue EPON 1004F, Miller-Stephenson or UV-curing glue Norland 81) onto a cylindrical fused silica disk (SurForce LLC) with the silver side towards the glue layer. In the section on Newton's rings force measurements, the surfaces were commercially purchased optical polished N-BK7 glass: a plano-convex lens with radius $R \approx 2$ cm (Edmund optics, #63-476) installed on the force measuring spring and an ultra-thin window (Edmund optics, #66-187) glued to the bottom port of the µSFA. Both surfaces were cleaned with piranha solution before experiments. Water used for the experiments was deionized (18.2 MΩ·cm with 2 ppb impurity counts) using an Integral 5 MilliQ filtration system. The capabilities of the µSFA were tested with several solutions: (1) Polystyrene beads of 4 µm in diameter with Texas-Red fluorophore (excitation 589 nm/ fluoresce 615 nm) were purchased from Molecular Probes (FluoSphere with sulfate, F8858). The original solution was diluted ten-fold in water and then injected between two mica surfaces. (2) A suspension of silica nanoparticles (silica Ludox® HS40) at 0.3 %w/w and of hydrodynamic diameter of 40 nm were functionalized with a fluorescent dye. (3) A series of mixtures of glycerol (Sigma-Aldrich) and water (glycerol volume concentrations of 0 %, 5 %, 10 %, 15 %, 20 %, and 25 %) was prepared before injecting 50 µL of each of these mixtures between mica surfaces in the µSFA. (4) A lysozyme solution of 1 %w/w at pH 4.5 was prepared in a phosphate-buffered saline (PBS, Sigma-Aldrich). A mica surface was incubated in the lysozyme solution for 1 hour and then rinsed and immersed in PBS. (5) A series of potassium chloride (KCl, Sigma-Aldrich) solutions of concentrations of 1 mM, 10 mM, and 100 mM was used to test the standard electrostatic double layer theory. All measurements were performed at 22°C.

**Surface Forces Apparatus (SFA) technique.** The Surface Forces Apparatus uses white light multiple beam interferometry to determine the separation distance and refractive index of thin films of vapor or liquid between two surfaces. A typical setup in the SFA involves symmetric mica surfaces (typically 1-5 µm thick mica sheets) back-coated with reflecting silver.[16] The interference pattern generated by monochromatic light passing through the mica surfaces in close proximity is known as Newton's rings. A light source emitting a continuous range of wavelength (e.g., white light) passing through the mica surfaces will form an interference pattern that, when passed through a spectrometer, appears as fringes of equal chromatic order (FECO). The absolute separation distance (as well as the separation distance profile along a line through the contact region) between the mica surfaces can be calculated from the FECO fringes.[22-27]

To measure a force-distance profile a piezo drive or a motorized micrometer drives the surfaces together (with a known distance displacement $\Delta D$ over a short time interval) from a region of no inter-surface force into a separation distance at which the surfaces interact, after which they are separated again. Over the same time interval, the actual distance displacement $\Delta D_a$ is calculated from the FECO. With a pre-calibrated spring of spring constant $k$, the inter-surface force can be calculated using Hooke's law: $F = k \cdot (\Delta D_a - \Delta D)$. A negative force corresponds to attraction and positive force corresponds to repulsion. The force $F$ between two curved surfaces with radius of curvature $R$ is converted to a pressure $P$ between two planar surfaces by using Derjaguin approximation to obtain the energy $E = F/(2\pi R)$, then differentiate to get $P = dE/dD$.[1] At flat contact the average pressure is $P = F/A$, where $A$ is the contact area.

**The miniature Surface Forces Apparatus (μSFA).** To make an SFA compatible with optical and spectroscopic techniques, we designed the μSFA shown in Figure 2. We (1) reduced the dimensions of the existing SFA (i.e., SFA2000 from SurForce LLC) by roughly a factor of two, i.e., reducing the volume by a factor of 8; (2) reconfigured the mechanical controls, electrical connections, fluid inlet and outlet ports, sample insertion port, thermistor port, viewing windows, etc., to enable operation of the μSFA during concurrent optical and spectroscopic measurements; (3) enlarged the entrance port or window to receive large fluorescence and other microscope objectives, (4) reduced the distance between the objective and sample from 10 mm to < 3 mm to match the focal length of typical fluorescence and other microscope objectives, and (5) enabled "normal" crossed-cylinder (equivalent to sphere-on-flat at short separation distances) and sphere-on-flat modes for spectroscopic and high-resolution microscopy (high numerical aperture and small working distance) applications. These design changes allow the μSFA to be mounted on a microscope stage and facilitate optical, fluorescence, and other types of spectroscopic imaging and characterization (e.g., Surface Plasmon Resonance and Raman) of systems under confinement.

The μSFA has four levels of positioning control spanning ranges of motion from mm to sub-nm: a differential micrometer with coarse and medium control, a motorized fine micrometer coupled to the lower surface via the force spring attachment, and a piezoelectric tube coupled to the top surface. The differential micrometer deflects the coarse displacement spring around a flexure point at the same plane as the contacting surfaces to ensure a vertical motion of the surfaces (see Fig. 2a, inset). A pair of anti-backlash springs (attached between the Attachment Base and the Main Chamber) counteract the motion of the differential micrometer and reduce mechanical noise. The fine micrometer deflects the fine displacement spring. The combined spring constant of the helical spring and the fine displacement spring is tuned to reduce the displacement of the lower surface 1000-fold with respect to the displacement of the fine micrometer. The surfaces (on glass disks) are mounted on dove-tailed disk mounts for easy insertion of samples.

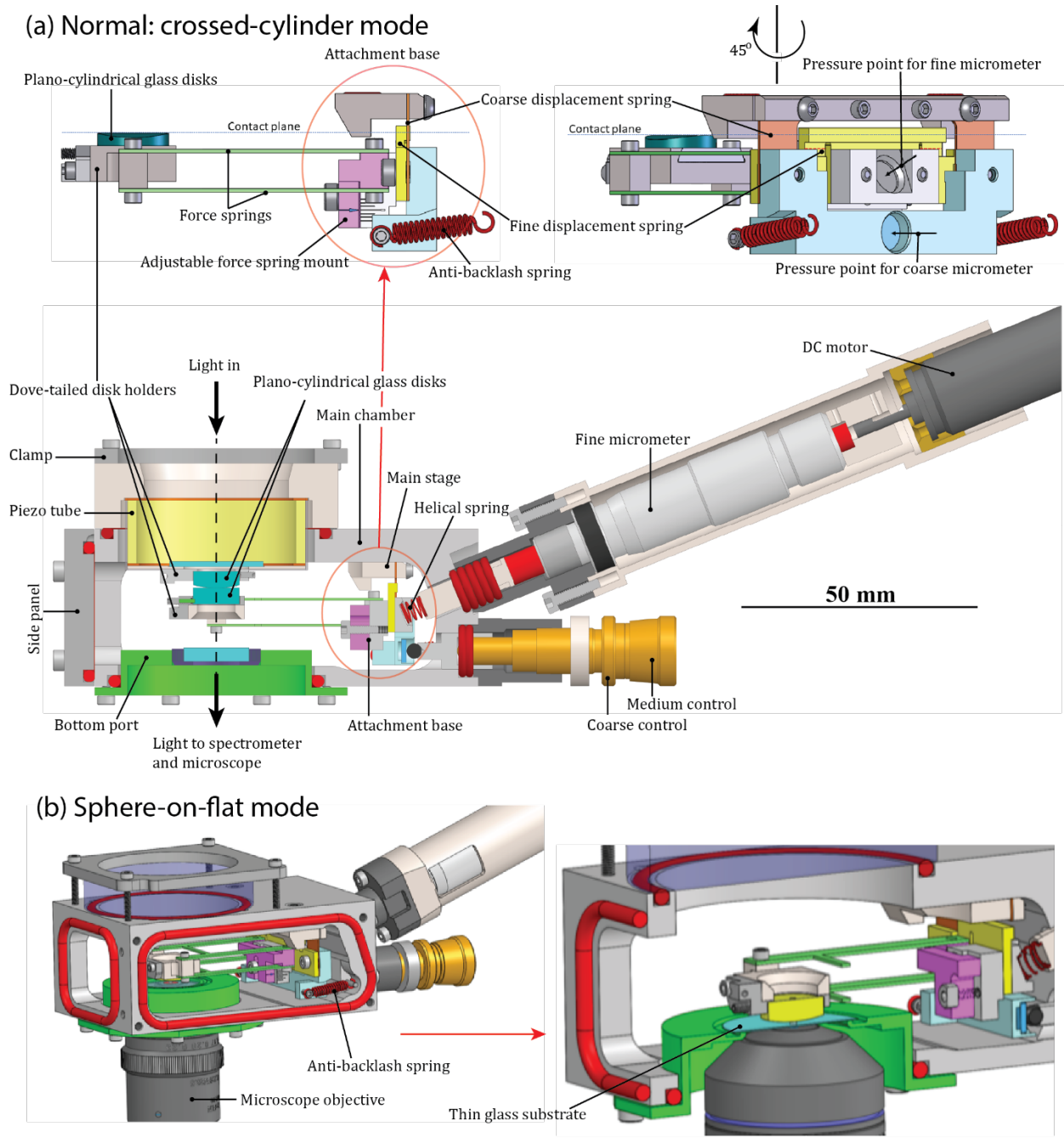

**Figure 2.** (a) Cross-section of the μSFA (with the back panel removed for clarity) with crossed-cylindrical glass disks. The inset shows the force spring attachment, and the top right shows the force spring attachment rotated by 45°. (b) Sphere-on-flat mode in the μSFA; here a spherical surface mounted on a double cantilever spring approaches downwards to a circular optical window
10

or cover slip of thickness 0.2 mm. The right panel shows a cross section view of how the μSFA accommodating a microscope objective, allowing the use of short working distance objectives.

The fluorescence microscope is an Olympus IX73 Dual Deck Microscope System with a 100W halogen optical white light source and a LED fluorescence light source. The microscope stage was custom-made to accommodate the μSFA (Fig. 1b).

The surface plasmon resonance (SPR) attachment consists of an input white light source (DH-2000-BAL, Ocean Optics), a collimator (74-UV, Ocean Optics), a polarizer (LPNIRE11S, Thorlabs), a pair of right-angle prisms to guide the light, a custom-made BK-7 glass prism with 45° angled walls that reflects the white light to the center of the SPR prism, an output collimator (74-UV, Ocean Optics) and an optical fiber (SPLIT-400-VIS-NIR, Ocean Optics) that is connected to the inputs of a spectrometer (HR4000, Ocean Optics).

As will be described in the following sections, the μSFA has been tested in both a standard SFA optical setup and an inverted microscope setup.

RESULTS AND DISCUSSION

The μSFA is designed to be integrated with new optical and spectroscopic attachments. First, however, we demonstrate that it is capable of reproducing standard SFA measurements.[28] Figure 3a shows normal forces, measured using the FECO technique described above, between two mica surfaces immersed in an aqueous solution of 5 mM NaCl. The forces agree well with the expected DLVO (van der Waals and electrostatic double layer) forces between the anionic mica surfaces. The inverted microscope allows facile imaging of the contact between mica surfaces. Figure 3b shows an image of the Newton's rings interference pattern that occur when bringing mica surfaces into contact in air. The inverted microscope also facilitates fluorescence imaging of the contact area between surfaces in the µSFA, as shown in Figure 3c, which shows 4 µm fluorescent beads confined between mica surfaces. The distinct circular shape is the outline of the contact area, as the beads have been pushed out of the contact to the edge of the contact area.

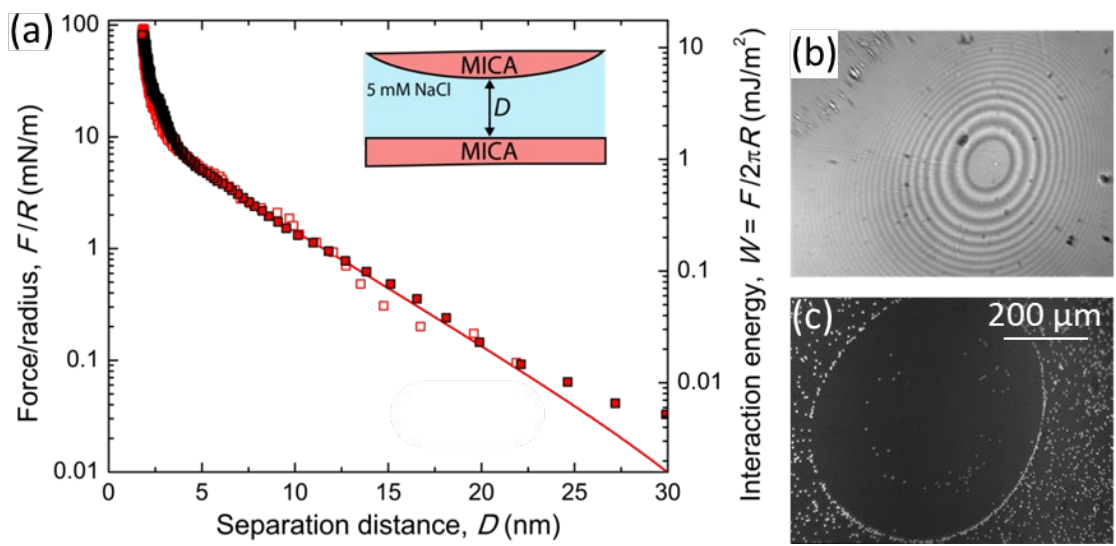

**Figure 3.** The µSFA has been utilized in (a) standard SFA normal force measurements between mica surfaces immersed in 5 mM NaCl. The red line is a theoretical fit including contributions from the expected electrostatic double layer repulsion, van der Waals attraction, and steric hydration repulsion.[1] The µSFA can also be used to image confined systems, demonstrated by (b) an image of the mica-mica contact Newton's rings, and (c) a fluorescence image of 4 µm diameter polystyrene beads confined between mica surfaces. Images in (b) and (c) were obtained with a 10× objective.

To demonstrate the new capabilities of the μSFA design, the following four sections describe experiments using fluorescence microscopy, Surface Plasmon Resonance (SPR) spectroscopy, and Raman spectroscopy in combination with normal (to the interacting surfaces) and frictional force measurements. The adaptability of the μSFA is showcased by these three examples, as each set of experiments uses a different surface configuration: crossed-cylinders (fluorescence), sphere-flat (SPR), and inverted sphere-flat (Raman).

**Simultaneous and In-Situ Force Measurements and Fluorescence Imaging (FL-μSFA)**

The new multimodal μSFA enables *in-situ* and simultaneous measurements of inter-surface forces and fluorescence imaging. Fluorescence microscopy has widespread applications in biological, polymer, and nanoparticle science.[29-33] Recently, a custom-designed fluorescence attachment to the SFA 2000 allowed visualization of the hemifusion of two lipid bilayers.[18] The μSFA may be mounted and operated in an inverted fluorescence microscope (Fig. 4a), allowing fluorescence excitation and imaging across a broad range of wavelengths. Fluorescent (FL) species between two SFA surfaces (Fig. 4b) are excited and imaged with the fluorescence microscope (Fig. 4c). Simultaneously, white light (with the wavelengths in the region of the fluorescence light filtered out) is passed between the surfaces. The resulting interference pattern is transmitted to a spectrometer producing FECO fringes (Fig. 4d). This FL-μSFA design allows imaging of fluorescently-tagged species such as biomembranes, nanoparticles, etc., under diverse conditions within the SFA, including under normal forces and/or shear. Furthermore, morphological changes or fluid dynamics can be tracked during force measurements.

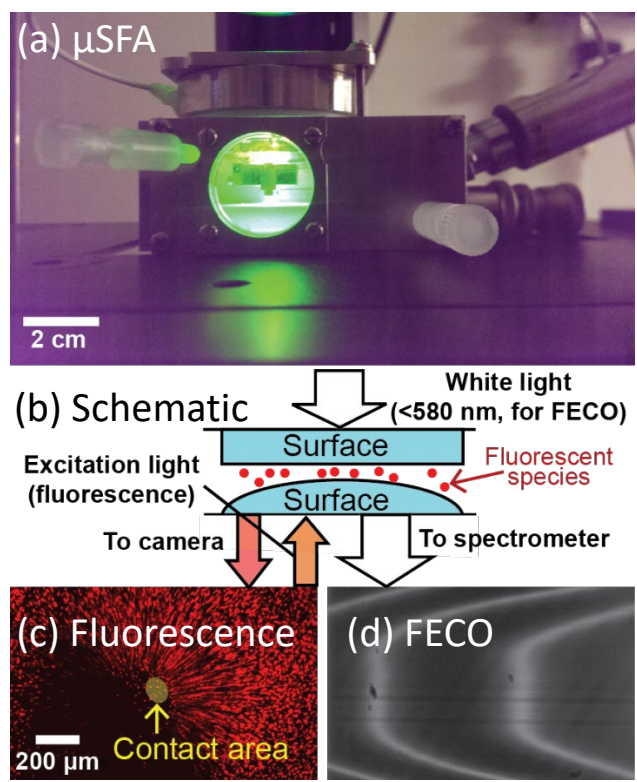

**Figure 4.** (a) The µSFA in an inverted microscope in fluorescence imaging mode. (b) Schematic illustration of a fluorescence µSFA experiment. Orange light (589 nm) is used to excite fluorescent species, which in this case emit at red wavelengths. Simultaneously, filtered white light (<580 nm) is reflected between the surfaces. The transmitted light near the point of contact creates Newton's rings (see Fig. 3b) for monochromatic light and fringes of equal chromatic order (FECO) for white light, as in a conventional SFA experiment. The combined methods enable *in-situ* (c) fluorescence imaging and force-distance measurements by analysis of (d) the FECO fringes. Panel (c) shows 4 µm diameter polystyrene beads with Texas-Red fluorophore confined between mica surfaces during approach and (d) shows the corresponding FECO.

To demonstrate the application of the new FL-µSFA setup for measuring particle and fluid dynamics, we analyzed forces between mica surfaces across a solution containing 2 mg/mL of 4 µm polystyrene beads tagged with Texas-Red fluorophore. Movements of the particles were tracked using the fluorescent microscope. Approximately 41,000 partial particle trajectories were detected and analyzed using MosaicSuite particle tracker[34] and Fiji[35]. A representative video of particle motions during approach along with the visualized particle trajectories are shown in the Supporting Information (Video S1).

The FL-µSFA setup allows simultaneous measurement and analyses of inter-surface pressures and the tracked particle motions (Fig. 5). The inter-surface pressure $P$ as a function time is shown in Figure 5a. The pressure starts to increase at 34 s as particles are trapped and compressed between the two surfaces. The pressure increases monotonically for the remainder of the approach, except between 37 and 41 s where a combination of rearrangement of the trapped microparticles and a lateral shift in the contact point of tens of µm results in a pressure drop. Concomitantly, the mean particle speeds were determined for particles within annular regions 50 µm in width of different distances from the center of the inter-surface contact area, which are shown in the inset to Figure 5b and in Figure S1. Within 50 µm of the center of the contact area (Fig. 5b, orange), the particles are relatively immobile during the first 25 s of the approach of the surfaces. However, further from the contact point the particles rapidly increase in speed 20 s after the beginning of the approach. The increase is most pronounced for particles 300-350 µm from the contact (Fig. 5b, green), and is less pronounced for particles closer (Fig. 5b, blue and magenta) and further (Fig. 5b, red) from the contact. As the inter-surface pressure increases for $t > 30$ s, the particles within the contact area (Fig. 5b, orange) slowly accelerate as the particles rearrange and the surfaces shift laterally under high compression, plateauing at speeds of 0.47 µm/s.

As shown for this model system of fluorescently-tagged microparticles, the FL-µSFA setup enables the positions, trajectories, and velocities of fluorescent species to be easily determined and analyzed during SFA force-distance measurements. This provides complementary information about dynamics and interactions across different length and time scales that was previously inaccessible. The newly-developed FL-µSFA technique is expected to provide valuable new insights into the interactions of organic and inorganic species, with simultaneous imaging of surfaces, particles, and interfaces.

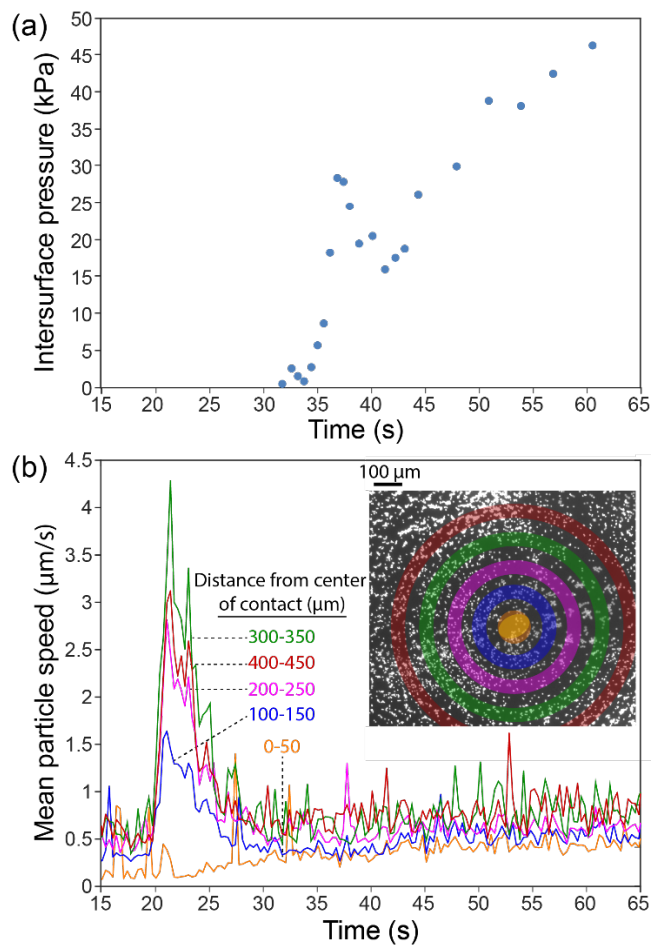

**Figure 5.** (a) Pressure between two mica surfaces across a suspension of 4 µm diameter polystyrene beads in the µSFA shown as a function of time as the surfaces are brought into contact.

Inter-surface pressure-distance and distance-time profiles are shown the Supporting Information (Fig. S2). (b) Mean particle velocities of 4 µm diameter polystyrene beads tagged with Texas-Red fluorophore confined between mica surfaces during approach within annular regions at selected distances from the center of the initial inter-surface contact: 0-50 µm (orange), 100-150 µm (blue), 200-250 µm (magenta), 300-350 µm (green), and 400-450 µm (red). The corresponding annular regions around the contact are shown in the inset FL image.

**Fluorescence Imaging and Tribological Measurements**

The field of tribology will particularly benefit from the combination of fluorescence imaging and surface forces measurements. Correlating frictional forces to the distribution, mobility, and conformations of lubricating molecules at interfaces has long been an elusive goal. The µSFA offers a unique opportunity to explore and unravel such correlations.

To demonstrate the capabilities of the µSFA to simultaneously measure lubrication forces and assess spatial distributions of lubricating species, we performed a series of tribological experiments to test the lubricating properties of suspensions of fluorescently labeled nanoparticles. To characterize the dynamic spatial distributions of the nanoparticles during shearing of the contact, we mounted the µSFA on a fluorescence microscope equipped with a camera for fluorescence imaging. The setup allows simultaneous acquisition of the FECO fringes emerging from the surfaces, frictional forces recorded by strain gauges installed on the upper surface holder and a top view image of the distribution of fluorescently labeled nanoparticles in the contact (see Fig. 4). To control the lateral motion of the upper surface, a standard SFA2000 friction device was used.[16]

This arrangement has proven to be particularly powerful to follow dynamically the initiation of wear cracks and their propagation along the shearing path. Identifying the initiation of wear during a tribological test is a long-standing challenge. The conventional way to detect wear initiation is through the monitoring of the frictional force during sliding or via analysis of the topography of the sliding track after experimentation. Monitoring the friction force to detect transient changes that could be correlated to surface damage is challenging and often subjective. Many systems that are prone to wear do not necessarily demonstrate any changes in friction coefficient[36] and, similarly, drastic changes in the friction forces during sliding do not always correlate with surface damage.[37] *Ex-situ* analyses of the contact allow quantification of the chemical and structural changes along the wear track but provide minimal information on how these changes evolve over time.

The configuration of the μSFA in a fluorescence microscope (Fig. 4) provides a unique solution to this problem. The FECO provides a live cross section of the contact geometry with angstrom resolution in the vertical axis and micron resolution in the plane of sliding. Such high resolution in the z axis allows the detection of minute changes in the boundary lubricant film thickness which often precede the build-up of local pressure and subsequent damage.[38] Since the FECO only represents a single cross section of the contact (either perpendicular to the sliding direction or parallel to it) unless scanning the SFA chamber through the optical path, wear initiation sites located far from the cross section are not visible in the FECO fringes. Concurrent analyses of FECO fringes with top view imaging of the entire contact is therefore advantageous. Top view imaging of the contact allows detection of changes in the refractive index of the confined lubricant film which could be related to changes in film thickness.[39-40]

To demonstrate the advantages of the μSFA with fluorescence imaging, we performed a tribological test using two mica surfaces immersed in a colloidal suspension of silica particles of hydrodynamic diameter $D_h$ = 40 nm. The particles were functionalized with a fluorescent dye to facilitate their detection via epifluorescence imaging. As can be seen in Figure 6a, the tribological results of a standard friction test measuring the friction force, $F_\parallel$, as a function of the normal load, $F_\perp$, show a linear dependence between these two quantities, a phenomenon described by Amonton's laws of friction.[1] The linear relationship between $F_\parallel$ and $F_\perp$ suggests that the friction coefficient between the two surfaces (defined as $\mu = F_\parallel/F_\perp$) is constant and independent of $F_\perp$ which would appear to indicate that surface damage did not occur during the test. Furthermore, as can be seen in Figure 6b, the FECO fringes do not necessarily present any sign of damage of the surface but rather local deformations. As previously reported, signs of surface damage appear in the FECO fringes as discontinuities in the fringe pattern due to loss of mica or even detachment of mica from the reflective layer.[41] However, top view fluorescence imaging of the contact reveals a very different situation. Real-time imaging shows accumulation of particles in multiple wear tracks parallel to the sliding direction and distributed throughout the contact, a phenomenon called the "mending effect".[42]

Observation of this phenomenon is almost impossible with FECO alone or bright field imaging of the top view. Fluorescence imaging (Fig. 6c) demonstrates that surface damage does occur, and that crack filling by nanoparticles during shear almost instantaneously mitigates the expected fluctuations in friction forces normally observed during wear initiation and propagation, resulting in a linear relationship between frictional forces and load despite the surface damage.

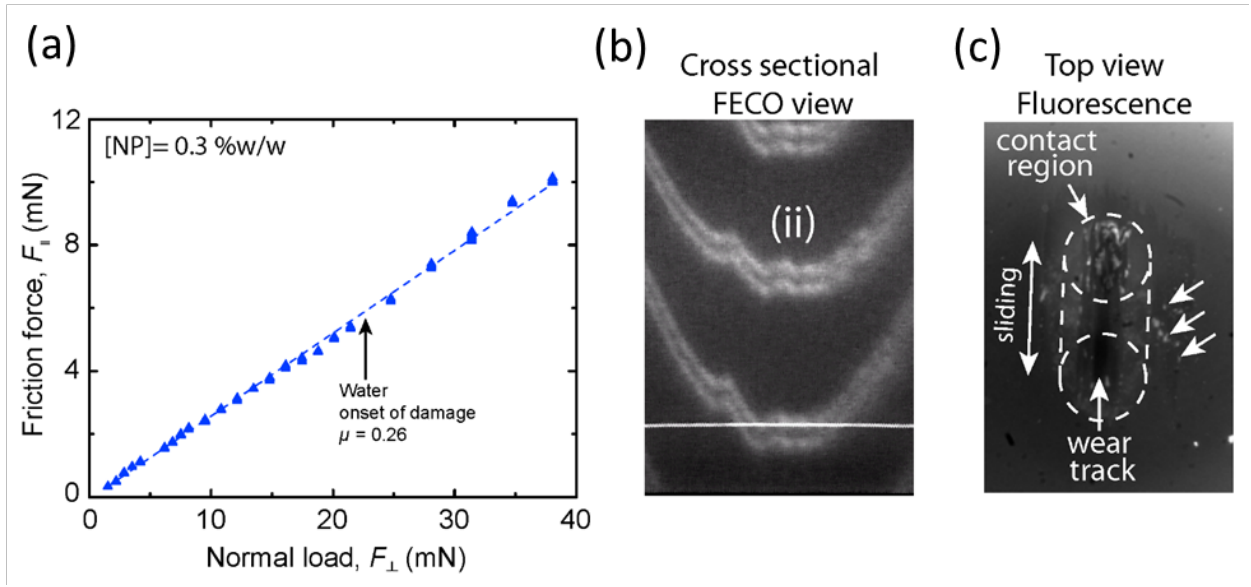

**Figure 6.** (a) Tribological test between two mica surfaces immersed in a suspension of silica nanoparticles at 0.3%w/w. The linear relationship between the normal force $F_\perp$ and the friction force $F_\parallel$ over the entire load ranges apparently indicates that no measurable damage of the surfaces occurred during the test. (b) FECO image of the contact obtained at high load demonstrating local deformation of the contact due to uneven distribution of the nanoparticles in the contact area. The view does not show any signs of contact damage. (c) Top view of the contact obtained by fluorescence imaging of the silica nanoparticles. The image shows the sliding trajectory of the contact with its contour as well as regions depleted in nanoparticles which are heavily damaged. Particle aggregates produced during shearing can be seen outside of the contact area (white arrows).

**Surface Plasmon Resonance and Surface Forces (SPR-μSFA)**

Surface plasmon resonance (SPR) spectroscopy is a useful tool for characterizing and quantifying (bio)molecular interactions and measuring the binding of analytes to immobilized biomolecules without using labels.[43] The most common SPR sensors include a gold or silver film (40-60 nm thickness)[44] deposited on a prism. This film thickness is similar to the gold or silver layers (40-55nm thickness) used as reflecting layer in the interferometric technique in the SFA. Therefore, it is possible to design an SPR attachment for the SFA, as seen in Figure S3, that allows simultaneous detection of refractive index changes near the SPR surface (e.g., from molecular binding and mass transfer) while performing SFA measurements.

Figure 7a depicts the setup for detection of SPR within an SFA. For the SPR attachment, light from a white light source is passed through a collimator to produce a parallel beam. After passing the polarizer, p-polarized incident light is produced that propagates towards a glass prism coated with a gold layer (Fig. 8b). Under the condition of attenuated total reflection (ATR), the energy of incident light is coupled to excite surface plasmons and consequently weaken the intensity of reflected light.[45] The reflected light is coupled to an output collimator and guided through an optical fiber and into a spectrometer. As the resonant wavelength is sensitive to the surface refractive index (less than 500 nm thick layer from the surface), and the surface refractive index typically changes upon molecular adsorption, we can monitor the molecular binding behavior on the gold surface through the changes to the SPR resonant wavelength.[46] This setup allows for simultaneous SFA and SPR measurements.

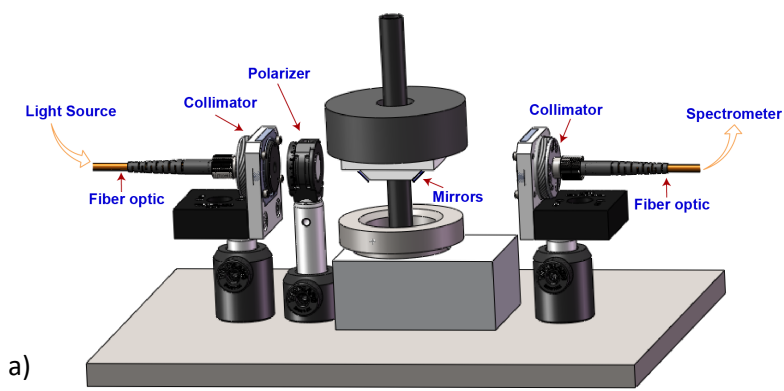

a)

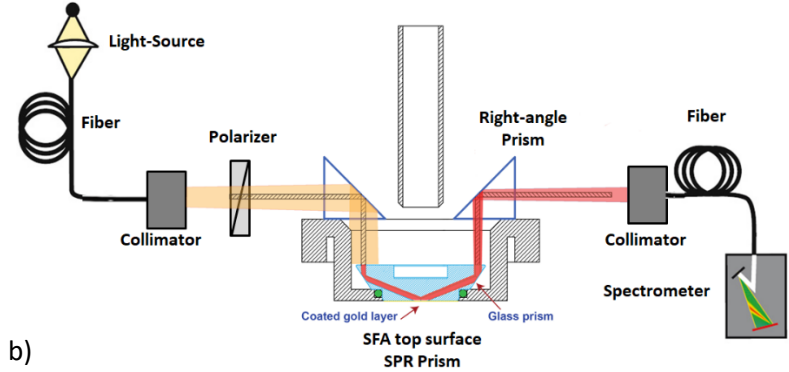

b)

**Figure 7.** (a) Sketch of the SPR-µSFA setup, and (b) schematic of a glass prism coated with a gold layer sitting in the SFA top mount. A polarized input white light is guided through an optical fiber and glass prisms to the center of the SPR prism (top surface of the SFA). The resulting surface plasmon resonance (SPR) signal is guided through an optical fiber and glass prisms to a spectrometer connected to a computer for data collection. The SPR surface is a mica-templated layer of 43 nm gold. The SPR attachment is coupled with a µSFA setup.

In order to evaluate the sensitivity of the SPR-µSFA to the surface refractive index, the resonant wavelengths are measured for various water/glycerol mixtures (v/v% glycerol = 0-25%) with refractive index (RI) between 1.333 and 1.368. Figure 8a shows the increase (i.e., redshift) in SPR wavelength as the concentration of glycerol in the water/glycerol mixtures increases. Figure 8b shows the relationship between SPR wavelength and the refractive index, which confirms the refractive index sensitivity of the SPR wavelength.

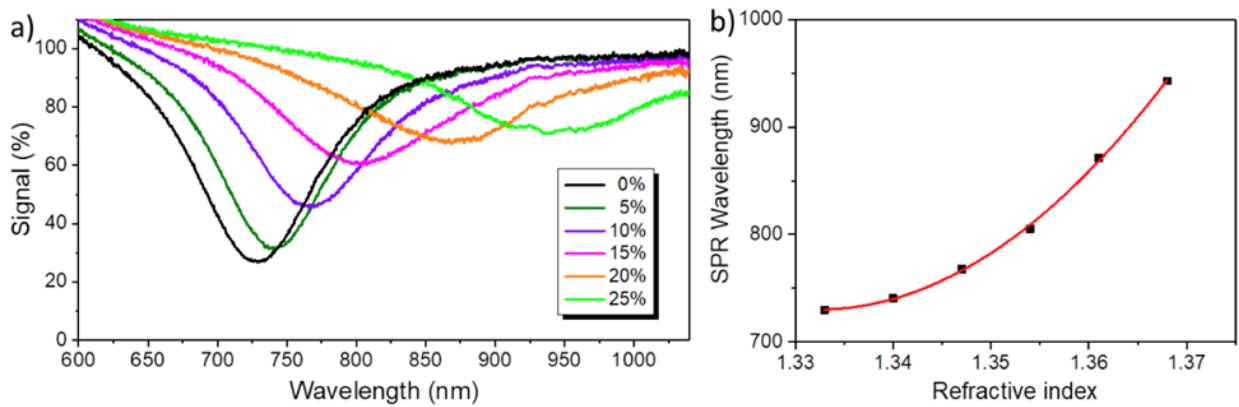

**Figure 8.** (a) Reflectivity spectra of the SPR-μSFA setup in water/glycerol mixtures with different concentrations of glycerol (v/v% glycerol = 0-25 %). (b) Plot of the resonant wavelength shift vs. the refractive index (RI) value measured with the SPR-μSFA. The red curve is a fit to the data.

To demonstrate the ability of the SPR-μSFA to detect protein mass transfer, we measured the mass transfer of lysozymes from the apposing mica surface to the gold surface. The mica surface was incubated for 1 hour in a 1 %w/w lysozyme solution (pH=4.5) allowing lysozyme to physically adsorb onto the mica surface. The mica was then rinsed with PBS, resulting in a lysozyme film on the mica as depicted schematically in Figure 9a. The lysozyme film was brought into contact with the gold surface, held for 3 min, and then separated. The approach and separation velocities were 5 nm/min. The reflectivity spectra and corresponding SPR wavelength values were recorded before and after each contact. Figure 9b shows that after the first contact of the two surfaces, lysozyme molecules were partially transferred from mica surface to gold surface, resulting in an increase in SPR wavelength. After the second approach and separation, there was no further increase in SPR wavelength, which indicated the mass transfer of lysozyme from mica to gold mainly occurred during the first contact. Further contact between the two surfaces did not induce any mass transfer.

The surface forces can provide information about changes in adhesion and other inter-surface interactions, and will be the subject of more detailed analyses in the future.

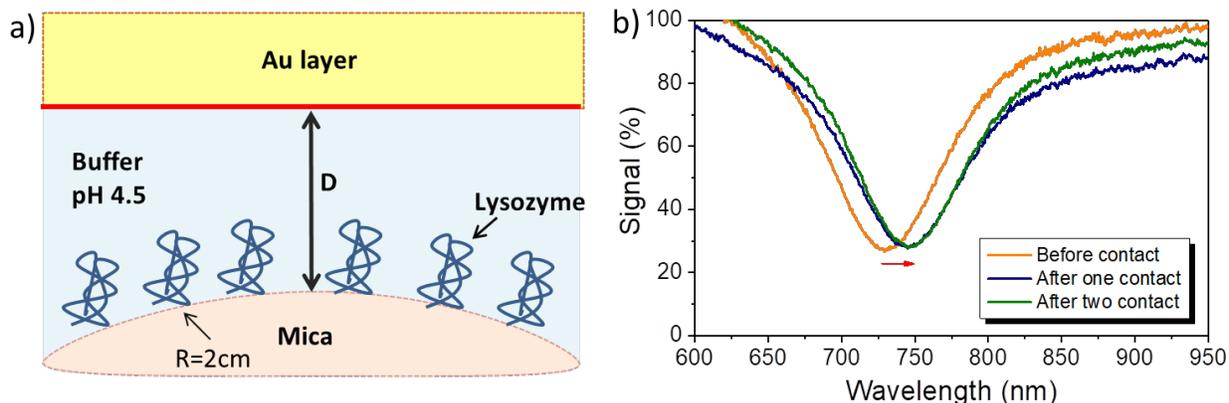

**Figure 9.** (a) Schematic of the interaction between a mica substrate coated with lysozyme and a bare gold (Au) top surface. (b) Reflectivity spectra of the SPR-μSFA setup before and after the two surfaces come into contact.

The SPR-μSFA setup also enables real-time monitoring the kinetics of specific binding as a function of the separation distance and to *in situ* examination of the changes in the conformations of molecules (e.g., polymers) adsorbed on or trapped between the surfaces under normal and shear forces. With the aid of the accurate distance control/measurement of the μSFA, the effects of distance between the gold nanofilm and adsorbed molecules (at the apposing surface) on the SPR signal can be directly studied.

**Newton's rings-μSFA and Raman-μSFA**

To make the μSFA compatible with microscopy and spectroscopy techniques that require short working distance (i.e., 100's of micrometers) between objective lens and sample or special sample substrates, an optical analysis technique based on Newton's rings was implemented to capture high-resolution force-distance data between glass surfaces. Here, we describe this Newton's rings method, and demonstrate that the method allows the μSFA to be coupled with Raman spectroscopy.

As discussed earlier, the classical SFA uses back-silvered mica in a crossed-cylinder geometry to obtain FECO fringes to measure surface separation distance,[16, 28] which results in several limitations. First, mica is notoriously difficult to prepare with the proper degree of cleanliness.[21, 47] Mica is also mostly inert and can only be functionalized after plasma activation which provides a higher density of reactive surface silanols.[48] The back-silvering results in ~95% reflectance, which makes optical measurements difficult. The loss of intensity due to this high reflectance has been overcome by using dielectric coatings with engineered wavelength transmission windows to enable spectroscopic techniques (e.g., fluorescence microscopy),[18, 49-50] but mica surfaces with such coatings are difficult to prepare. In spectroscopic applications, the birefringence of mica distorts the point-spread-function of the microscope and makes reproducible spectroscopic analyses difficult, particularly for Raman spectroscopy.[51-52] Finally, the mica is typically mounted on cylindrical silica lenses with thicknesses of several mm.[16] These lenses therefore only accommodate microscope objectives with a working distance of generally ≥ 10 mm. For these reasons, there are few examples of combining spectroscopy with force measurement or controlled confinement in the literature,[53-55] none of which have reached widespread use to date.

To overcome the limitations of short working distance of the objective lens and back-silvered mica substrates, we present here a sphere-on-flat measurement mode for μSFA that uses high-resolution distance measurements by analysis of Newton's rings observed between glass surfaces. We henceforth refer this method as Newton's rings-SFA (NR-μSFA). As shown in Figure 2b, the μSFA can be configured such that a spherical glass lens mounted on the double cantilever spring approaches a flat glass coverslip glued on the bottom port of the μSFA. The μSFA is designed to accommodate the microscope objective (Fig. 2b) such that the objective can reach the bottom of the coverslip, permitting the use of very short working distance objectives with high numerical aperture in epi-illumination, enabling high-resolution imaging. This configuration also eliminates many of the limitations of preparing mica surfaces by using optically transparent glass surfaces, which are easy to clean and functionalize (however, with the trade-off that they are not atomically smooth as in the case of mica).

A simplified schematic of the geometry is shown in Figure 10a. The contact area is illuminated with monochromatic light of wavelength $\lambda$ and the light interferes between the two surfaces of refractive index $n_2$ through the medium refractive index $n_1$. The reflected light is imaged in an inverted microscope to acquire Newton's Rings patterns during the approach and separation of the top sphere. An example of a Newton's Rings pattern is shown in Figure 10b. This is a combined image of a raw Newton's Rings image divided by a background reference image. The reference image is taken at a large surface separation where no observable radial intensity pattern is found. The Newton's rings pattern is then radially averaged around its center point to obtain the intensity profile as a function of radial distance $I(r)$ (Fig. 10c).

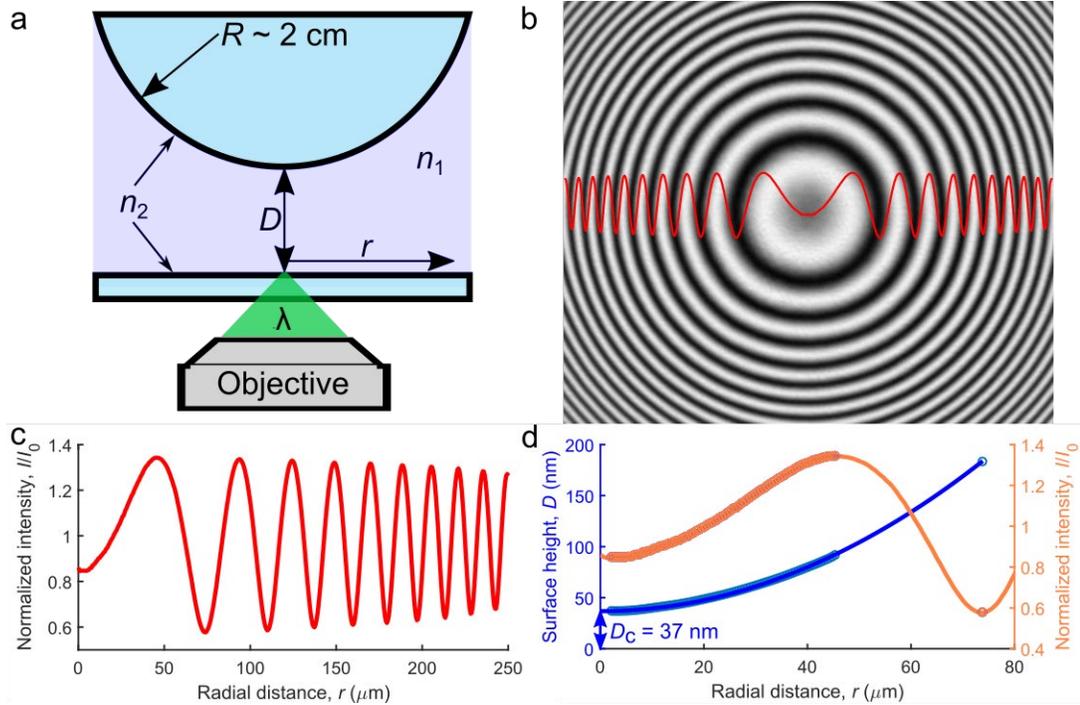

**Figure 10.** Newton's rings method for measuring separation distance. (a) A schematic of the sphere-on-flat configuration illuminated by monochromatic light of wavelength λ that gives (b) a high-resolution Newton's rings pattern. The Newton's rings in (b) are radially averaged to obtain (c) the intensity profile $I(r)$ which can be analyzed as discussed in the text to reconstruct the (d) surface profile $D(r)$ and the center distance $D_C$.

The shape of the surface can be reconstructed from $I(r)$, a procedure commonly used in Reflection Interference Contrast Microscopy (RICM).[56-60] Newton's rings have been used for force measurements as far back as Derjaguin in 1958,[61] and RICM itself has been used to measure surface forces of air bubbles and other soft particles previously.[57-58] A similar RICM shape-reconstruction procedure was used recently for AFM force measurements.[59] Here, we calculate the surface profile $D(r)$ from $I(r)$ using Equation 1, which accounts for the intensity profile within the first extremum,

$$D(r) = \frac{\lambda}{4\pi n_1} \cos^{-1}\left(\frac{A-I(r)}{B}\right), \tag{1}$$

where $A = (I_{max} + I_{min})/2$, and $B = (I_{max} - I_{min})/2$. For $r$ outside of the first extremum, a height increment of $\lambda m/4n_1$ is used to calculate $D(r)$ at each extremum, where $m$ is the extremum number. Equation 1 assumes parallel planar interfaces, an appropriate assumption given the large radius of curvature of surfaces commonly used in SFA, typically $R \sim$ 1-2 cm.[59] Deviations from parallel surfaces are at maximum $dD/dr \sim 10^{-3}$ for the lateral length scales encountered here.

An example of the intensity profile measurement and corresponding surface profile calculation is shown in Figure 10d for the first two extrema. The intensity profile (orange curve, Fig. 10d) at small values of $r$ contains a small number of pixels included in the radial average and is generally excluded from the analysis. The height profile (blue points, Fig. 10d) is calculated from the intensity profile (orange points, Fig. 10d) using Equation 1 and the first two extrema. In this example, a spherical shape ($R$ = 1.9 cm) fits the intensity data, (blue curve, Fig. 10d) and the fit yields the separation distance at the center of the contact area, i.e., $D(r=0) = D_C$ = 37 nm, which is the relevant distance for the force measurement. Due to the geometry of the surfaces, nm-scale movement of the surface corresponds to micron-scale movement of $I(r)$ in the region $r \sim$ 5-80 μm, which is easily resolvable with conventional optical microscopy. Therefore, this surface reconstruction procedure enables measurement of $D_C$ with nm-level resolution, allowing for accurate measurement of force-distance profiles, as shown further below.

To demonstrate the utility and accuracy of NR-μSFA, we measured normal force as function of separation distance between glass surfaces in 1 mM, 10 mM, and 100 mM KCl, as shown in Figure 11a. The long-range interactions are well-fitted by standard electrostatic double layer theory, and follow an exponential force-distance relation with typical decay length, $\kappa^{-1}$ known as the Debye length, that depends on the solution's ionic strength.[1] Importantly, the fitted Debye length corresponds closely to the expected Debye length at each salt concentration. We measured decay lengths of 8 nm, 4 nm, and 1 nm (Fig. 11a) for theoretical Debye lengths $\kappa^{-1}$ of 9.6 nm, 3.9 nm, and 0.96 nm, respectively. Repeated measurements overlay each other in the force-distance profile, demonstrating that the distance measurement is reliable and the interactions are reversible (i.e., the same on approach and retraction) and reproducible. To further quantify the distance accuracy, the inset of Figure 11a shows a zoomed-in view of a single force profile at high force. The observed noise envelope of ~0.5 nm is due to slight variability in the optical analysis and sphere fitting procedure. This noise level is satisfactory for measuring most colloidal interactions, especially long-range interactions.

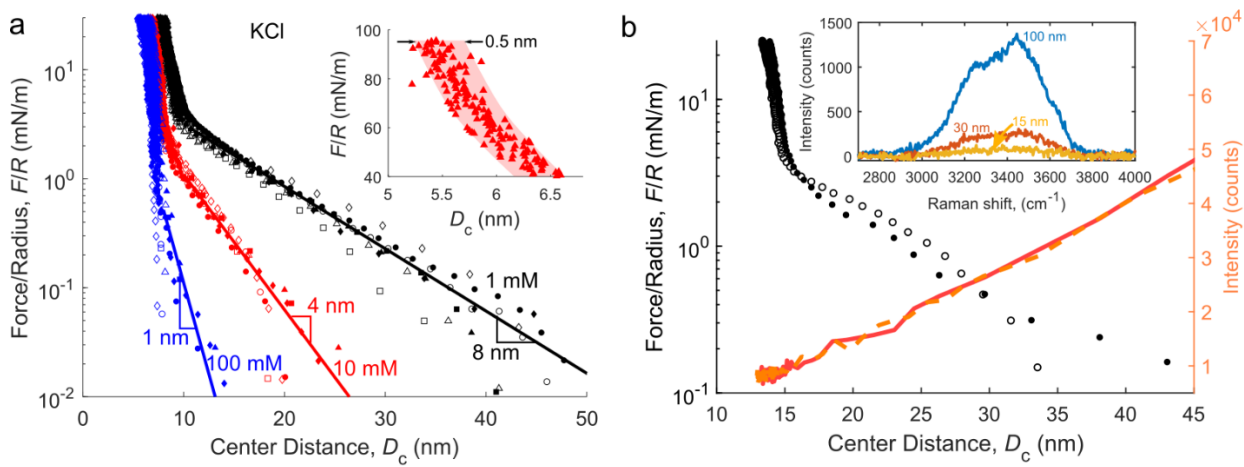

**Figure 11.** (a) Force profiles measured by NR-μSFA in 1 mM KCl (black), 10 mM KCl (red), and 100 mM KCl (blue). Closed points are measured during approach and open points are measured during separation. The different symbols represent separate approach-retraction measurements (4 sets of approach-retract are displayed for each salt concentration). The lines represent a best fit of the electrostatic forces and the measured decay length is indicated for each salt concentration. The inset shows a single run at high force to illustrate the relative noise level of the distance measurement. (b) Force-distance profile (left axis: black points, filled points-approach and open points-separation) and correlated Raman signal intensity (right axis: orange lines, solid and dashed corresponding to approach and separation, respectively) of the water between the surfaces measured simultaneously as described in the text. The inset shows the Raman spectra in the OH stretch region at different values of $D_C$.

To demonstrate the potential of NR-μSFA for multi-modality, especially in applications that require high-resolution imaging capabilities, we incorporated the μSFA into a home-built inverted microscope to measure force-distance curves simultaneously with Raman spectroscopy (we note that the NR-μSFA can also be integrated into commercially available confocal microscopes). The design and operation of the Raman-μSFA will be fully detailed in a forthcoming publication. Briefly, the design is based on a conventional confocal Raman microscope, and relies on two independent light paths for simultaneously obtaining Newton's rings and Raman spectra with the same high NA objective. A green laser is used for Raman excitation while the Newton's rings are imaged with blue light; standard dichroic mirrors are used to separate the light paths and steer the Raman signal into a spectrometer.

To demonstrate concurrent Raman-μSFA spectroscopy measurements, we simultaneously measured force-distance profiles by Newton's rings and Raman spectra during approach and separation of two glass surfaces immersed in 1 mM KCl, as shown in Figure 11b. A Raman spectrum was acquired synchronously with the Newton's rings images taken at 1 frame per second, and the surfaces velocity on approach and separation was 6 nm/s in the zero-force regime. The force curve obtained from NR-μSFA is nearly identical to the one shown for 1 mM KCl in Figure 11a. The measured decay length of ~9 nm is close to the theoretical Debye length $\kappa^{-1}$ = 9.6 nm in 1 mM KCl predicted by Debye-Hückel theory.[1] The inset of Figure 11b displays background-subtracted Raman spectra at different surface separation distances. The broad O-H stretch resonance observed in this spectral region is assigned to water confined between the two glass surfaces. Integrating the Raman signal intensity over this 3100-3700 cm$^{-1}$ spectral region of the O-H stretch, and plotting this intensity as a function of the measured distance (Fig. 11b), shows a linear decrease in intensity as the surfaces are confined. The linear behavior of the Raman intensity with separation distance results from two aspects: (i) the vibrational density of states does not change upon confinement and (ii) the separation distances of interest ($D_C$ < 100 nm) are much smaller than the longitudinal resolution of the 532-nm focus. These results show that we can measure correlated changes in the Raman spectrum during approach and retraction of the surfaces in the μSFA. This multimodal Raman-μSFA technique is therefore useful for probing chemical information near interfaces during confinement.

It is interesting to note that the minimum separation distance (hard-wall thickness) of the surfaces was $D_C$~13 nm. The measurement shown in Figure 11a exhibits a similar nano-scale hard-wall at $D_C$~5 nm. The discrepancy between the hard-wall thicknesses measured in these experiments is likely due to heterogeneous asperities on the glass surfaces. The optical grade glass surfaces used here are polished but exhibit nanometer-level roughness, as measured by atomic force microscopy (AFM) shown in Figure S4. The overall root-mean-square surface roughness is 0.6 nm, but several asperities ranging from 5-15 nm in height are observed over a representative 20×20 μm area. Provided large asperities do not arise in the center of the contact zone, representative force measurements can be made to obtain results shown in Figure 11. The change in slope observed at $D_C$~10 nm in Figure 11a and $D_C$~15 nm in Figure 11b is likely due to compression of such asperities.

As presented here, there are both advantages and disadvantages of NR-µSFA compared to the SFA technique involving FECO fringes. Sample preparation is greatly simplified, it is easily adaptable for multimodal techniques, can be implemented on most standard inverted microscopes without additional required equipment, short working distance objectives can be used, and sub-nm distance resolution can be achieved. As with the standard SFA technique, particle contamination is sometimes unavoidable, and in this case can be detected via the Newton's rings by direct observation of microscopic particles, or if the surfaces exhibit unexpected repulsive forces. The absolute distance measurement is less direct than FECO and it is necessary to confirm that one is measuring the contact fringe by applying high force at the end of an experiment. More rigorous optical modeling of the Fresnel coefficients is currently underway and should provide more direct distance information. While the measurements shown here demonstrate that clean contact zones can be observed with commercially purchased surfaces, the roughness of the glass surfaces is not ideal for molecular scale force measurements; future efforts will include preparation of smoother surfaces using recent methods.[62] Nonetheless, we have shown here that NR-µSFA enables straightforward sample preparation, accurate force measurements, and facile multi-modality, including Raman spectroscopy.

CONCLUSIONS

We have shown that standard force-distance measurements are as feasible and accurate with the new µSFA as measurements performed with the established standard Surface Forces Apparatus, the SFA2000. Furthermore, the µSFA has the additional benefit of enabling concurrent optical and spectroscopic measurements of systems under confinement and shear. Here we have shown the feasibility of concurrent SFA measurements with fluorescence microscopy, Surface Plasmon Resonance Spectroscopy, and Raman Spectroscopy, offering promise that additional techniques can also be incorporated into the μSFA due to its small size and large ports. The benefit of simultaneous measurements is to unambiguously establish relationships between quantities measured using complementary techniques under identical conditions (i.e., same temperature, salt concentration, pH, and time-dependent history of the surfaces and liquid). The complementary multimodal measurements enabled by the μSFA are expected to provide essential insight into diverse, complex, and sensitive systems with broad relevance for understanding tribological, biological, and physiochemical phenomena at interfaces.

AUTHOR INFORMATION

**Corresponding Author**

*Email: kai_kristiansen@ucsb.edu; Phone: +1-805-893-5268.

**Present Addresses**

ϑCurrent address (Zachariah J. Berkson): Department of Chemistry and Applied Sciences, ETH Zurich, 8093 Zürich, Switzerland.

ACKNOWLEDGMENT

The development of the μSFA has been supported by the US Department of Energy, Office of Basic Energy Research, Division of Materials Sciences and Engineering under Award #DE-FG02-87ER45331 and by SurForce LLC (JS). SHD, HBA, and JDM were supported by LabEX ENS-ICFP: ANR-10-LABX-0010/ANR-10-IDEX-0001-02 PSL*. GDD was supported by the National Science Foundation Graduate Research Fellowship Program under Grant No. 1650114.

# REFERENCES

1. Israelachvili, J. N., Intermolecular and Surface Forces, 3rd Edition. *Intermolecular and Surface Forces, 3rd Edition* **2011**, 1-674.
2. Duke, C. B., The birth and evolution of surface science: Child of the union of science and technology. *P Natl Acad Sci USA* **2003,** *100* (7), 3858-3864.
3. Boles, M. A.; Ling, D.; Hyeon, T.; Talapin, D. V., The surface science of nanocrystals. *Nat Mater* **2016,** *15* (2), 141-153.
4. Kim, B. Y. S.; Rutka, J. T.; Chan, W. C. W., Current Concepts: Nanomedicine. *New Engl J Med* **2010,** *363* (25), 2434-2443.
5. Papastavrou, G., Combining electrochemistry and direct force measurements: from the control of surface properties towards applications. *Colloid Polym Sci* **2010,** *288* (12-13), 1201-1214.
6. Ding, S. Y.; Yi, J.; Li, J. F.; Ren, B.; Wu, D. Y.; Panneerselvam, R.; Tian, Z. Q., Nanostructure-based plasmon-enhanced Raman spectroscopy for surface analysis of materials. *Nat Rev Mater* **2016,** *1* (6), 16021.
7. Wang, G.; Rodahl, M.; Edvardsson, M.; Svedhem, S.; Ohlsson, G.; Hook, F.; Kasemo, B., A combined reflectometry and quartz crystal microbalance with dissipation setup for surface interaction studies. *Rev Sci Instrum* **2008,** *79* (7), 075107.
8. Zhang, N.; Schweiss, R.; Zong, Y.; Knoll, W., Electrochemical surface plasmon spectroscopy - Recent developments and applications. *Electrochim Acta* **2007,** *52* (8), 2869-2875.
9. Kircher, M. F.; de la Zerda, A.; Jokerst, J. V.; Zavaleta, C. L.; Kempen, P. J.; Mittra, E.; Pitter, K.; Huang, R. M.; Campos, C.; Habte, F.; Sinclair, R.; Brennan, C. W.; Mellinghoff, I. K.; Holland, E. C.; Gambhir, S. S., A brain tumor molecular imaging strategy using a new triple-modality MRI-photoacoustic-Raman nanoparticle. *Nat Med* **2012,** *18* (5), 829-834.
10. Golan, Y.; Seitz, M.; Luo, C.; Martin-Herranz, A.; Yasa, M.; Li, Y. L.; Safinya, C. R.; Israelachvili, J., The x-ray surface forces apparatus for simultaneous x-ray diffraction and direct normal and lateral force measurements. *Rev Sci Instrum* **2002,** *73* (6), 2486-2488.
11. Valtiner, M.; Banquy, X.; Kristiansen, K.; Greene, G. W.; Israelachvili, J. N., The Electrochemical Surface Forces Apparatus: The Effect of Surface Roughness, Electrostatic Surface Potentials, and Anodic Oxide Growth on Interaction Forces, and Friction between Dissimilar Surfaces in Aqueous Solutions. *Langmuir* **2012,** *28* (36), 13080-13093.
12. Banquy, X.; Kristiansen, K.; Lee, D. W.; Israelachvili, J. N., Adhesion and hemifusion of cytoplasmic myelin lipid membranes are highly dependent on the lipid composition. *Bba-Biomembranes* **2012,** *1818* (3), 402-410.
13. Banquy, X.; Lee, D. W.; Kristiansen, K.; Gebbie, M. A.; Israelachvili, J. N., Interaction Forces between Supported Lipid Bilayers in the Presence of PEGylated Polymers. *Biomacromolecules* **2016,** *17* (1), 88-97.
14. Min, Y.; Kristiansen, K.; Boggs, J. M.; Husted, C.; Zasadzinski, J. A.; Israeiachvill, J., Interaction forces and adhesion of supported myelin lipid bilayers modulated by myelin basic protein. *P Natl Acad Sci USA* **2009,** *106* (9), 3154-3159.
15. Israelachvili, J. N.; Tabor, D., Measurement of Van-Der-Waals Dispersion Forces in Range 1.4 to 130 nm. *Nature-Phys Sci* **1972,** *236* (68), 106.
16. Israelachvili, J.; Min, Y.; Akbulut, M.; Alig, A.; Carver, G.; Greene, W.; Kristiansen, K.; Meyer, E.; Pesika, N.; Rosenberg, K.; Zeng, H., Recent advances in the surface forces apparatus (SFA) technique. *Rep Prog Phys* **2010,** *73* (3), 036601.
17. Frechette, J.; Vanderlick, T. K., Double layer forces over large potential ranges as measured in an electrochemical surface forces apparatus. *Langmuir* **2001,** *17* (24), 7620-7627.

18. Lee, D. W.; Kristiansen, K.; Donaldson, S. H.; Cadirov, N.; Banquy, X.; Israelachvili, J. N., Real-time intermembrane force measurements and imaging of lipid domain morphology during hemifusion. *Nat Commun* **2015,** *6*, 7238.
19. Kristiansen, K.; Banquy, X.; Zeng, H. B.; Charrault, E.; Giasson, S.; Israelachvili, J., Measurements of Anisotropic (Off-Axis) Friction-Induced Motion. *Adv Mater* **2012,** *24* (38), 5236-5241.
20. Villey, R.; Piednoir, A.; Sharma, P.; Cottin-Bizonne, C.; Cross, B.; Phaner-Goutorbe, M.; Charlaix, E., Capacitive detection of buried interfaces by a dynamic surface force apparatus. *Rev Sci Instrum* **2013,** *84* (8), 085113.
21. Israelachvili, J. N.; Alcantar, N. A.; Maeda, N.; Mates, T. E.; Ruths, M., Preparing contamination-free mica substrates for surface characterization, force measurements, and imaging. *Langmuir* **2004,** *20* (9), 3616-3622.
22. Heuberger, M., The extended surface forces apparatus. Part I. Fast spectral correlation interferometry. *Rev Sci Instrum* **2001,** *72* (3), 1700-1707.
23. Heuberger, M.; Luengo, G.; Israelachvili, J., Topographic information from multiple beam interferometry in the surface forces apparatus. *Langmuir* **1997,** *13* (14), 3839-3848.
24. Kienle, D. F.; Kuhl, T. L., Analyzing refractive index profiles of confined fluids by interferometry part II: Multilayer and asymmetric systems. *Anal Chim Acta* **2016,** *936*, 236-244.
25. Levins, J. M.; Vanderlick, T. K., Extended Spectral-Analysis of Multiple-Beam Interferometry - a Technique to Study Metallic-Films in the Surface Forces Apparatus. *Langmuir* **1994,** *10* (7), 2389-2394.
26. Schwenzfeier, K. A.; Erbe, A.; Bilotto, P.; Lengauer, M.; Merola, C.; Cheng, H. W.; Mears, L. L. E.; Valtiner, M., Optimizing multiple beam interferometry in the surface forces apparatus: Novel optics, reflection mode modeling, metal layer thicknesses, birefringence, and rotation of anisotropic layers. *Rev Sci Instrum* **2019,** *90*, 043908.
27. Zappone, B.; Zheng, W. C.; Perkin, S., Multiple-beam optical interferometry of anisotropic soft materials nanoconfined with the surface force apparatus. *Rev Sci Instrum* **2018,** *89* (8), 085112.
28. Israelachvili, J. N.; Adams, G. E., Measurement of Forces between 2 Mica Surfaces in Aqueous-Electrolyte Solutions in Range 0-100 Nm. *J Chem Soc Farad T 1* **1978,** *74*, 975-1001.
29. Huang, B.; Bates, M.; Zhuang, X. W., Super-Resolution Fluorescence Microscopy. *Annu Rev Biochem* **2009,** *78*, 993-1016.
30. Ruedas-Rama, M. J.; Walters, J. D.; Orte, A.; Hall, E. A. H., Fluorescent nanoparticles for intracellular sensing: A review. *Anal Chim Acta* **2012,** *751*, 1-23.
31. Dempsey, G. T.; Vaughan, J. C.; Chen, K. H.; Bates, M.; Zhuang, X. W., Evaluation of fluorophores for optimal performance in localization-based super-resolution imaging. *Nat Methods* **2011,** *8* (12), 1027-1036.
32. Hama, H.; Kurokawa, H.; Kawano, H.; Ando, R.; Shimogori, T.; Noda, H.; Fukami, K.; Sakaue-Sawano, A.; Miyawaki, A., Scale: a chemical approach for fluorescence imaging and reconstruction of transparent mouse brain. *Nat Neurosci* **2011,** *14* (11), 1481-1488.
33. Okabe, K.; Inada, N.; Gota, C.; Harada, Y.; Funatsu, T.; Uchiyama, S., Intracellular temperature mapping with a fluorescent polymeric thermometer and fluorescence lifetime imaging microscopy. *Nat Commun* **2012,** *3*, 705.
34. Sbalzarini, I. F.; Koumoutsakos, P., Feature point tracking and trajectory analysis for video imaging in cell biology. *J Struct Biol* **2005,** *151* (2), 182-195.
35. Schindelin, J.; Arganda-Carreras, I.; Frise, E.; Kaynig, V.; Longair, M.; Pietzsch, T.; Preibisch, S.; Rueden, C.; Saalfeld, S.; Schmid, B.; Tinevez, J. Y.; White, D. J.; Hartenstein, V.; Eliceiri, K.; Tomancak, P.; Cardona, A., Fiji: an open-source platform for biological-image analysis. *Nat Methods* **2012,** *9* (7), 676-682.
36. Lee, D. W.; Banquy, X.; Das, S.; Cadirov, N.; Jay, G.; Israelachvili, J., Effects of molecular weight of grafted hyaluronic acid on wear initiation. *Acta Biomater* **2014,** *10* (5), 1817-1823.

37. Lee, D. W.; Banquy, X.; Israelachvili, J. N., Stick-slip friction and wear of articular joints. *P Natl Acad Sci USA* **2013,** *110* (7), E567-E574.
38. Banquy, X.; Lee, D. W.; Das, S.; Hogan, J.; Israelachvili, J. N., Shear-Induced Aggregation of Mammalian Synovial Fluid Components under Boundary Lubrication Conditions. *Adv Funct Mater* **2014,** *24* (21), 3152-3161.
39. Bureau, L., Nonlinear Rheology of a Nanoconfined Simple Fluid. *Phys Rev Lett* **2010,** *104* (21), 218302.
40. Bureau, L.; Arvengas, A., Drainage of a nanoconfined simple fluid: Rate effects on squeeze-out dynamics. *Phys Rev E* **2008,** *78* (6), 061501.
41. Homola, A. M.; Israelachvili, J. N.; Mcguiggan, P. M.; Gee, M. L., Fundamental Experimental Studies in Tribology - the Transition from Interfacial Friction of Undamaged Molecularly Smooth Surfaces to Normal Friction with Wear. *Wear* **1990,** *136* (1), 65-83.
42. Liu, G.; Li, X.; Qin, B.; Xing, D.; Guo, Y.; Fan, R., Investigation of the mending effect and mechanism of copper nano-particles on a tribologically stressed surface. *Tribol Lett* **2004,** *17* (4), 961-966.
43. Kabashin, A. V.; Evans, P.; Pastkovsky, S.; Hendren, W.; Wurtz, G. A.; Atkinson, R.; Pollard, R.; Podolskiy, V. A.; Zayats, A. V., Plasmonic nanorod metamaterials for biosensing. *Nat Mater* **2009,** *8* (11), 867-871.
44. Subramanian, P.; Lesniewski, A.; Kaminska, I.; Vlandas, A.; Vasilescu, A.; Niedziolka-Jonsson, J.; Pichonat, E.; Happy, H.; Boukherroub, R.; Szunerits, S., Lysozyme detection on aptamer functionalized graphene-coated SPR interfaces. *Biosens Bioelectron* **2013,** *50*, 239-243.
45. Zeng, S. W.; Baillargeat, D.; Ho, H. P.; Yong, K. T., Nanomaterials enhanced surface plasmon resonance for biological and chemical sensing applications. *Chem Soc Rev* **2014,** *43* (10), 3426-3452.
46. Tokel, O.; Inci, F.; Demirci, U., Advances in Plasmonic Technologies for Point of Care Applications. *Chem Rev* **2014,** *114* (11), 5728-5752.
47. Lin, Z. Q.; Granick, S., Platinum nanoparticles at mica surfaces. *Langmuir* **2003,** *19* (17), 7061-7070.
48. Liberelle, B.; Banquy, X.; Giasson, S., Stability of silanols and grafted alkylsilane monolayers on plasma-activated mica surfaces. *Langmuir* **2008,** *24* (7), 3280-3288.
49. Mukhopadhyay, A.; Zhao, J.; Bae, S. C.; Granick, S., An integrated platform for surface forces measurements and fluorescence correlation spectroscopy. *Rev Sci Instrum* **2003,** *74* (6), 3067-3072.
50. Wang, Y. J.; Li, F.; Rodriguez, N.; Lafosse, X.; Gourier, C.; Perez, E.; Pincet, F., Snapshot of sequential SNARE assembling states between membranes shows that N-terminal transient assembly initializes fusion. *P Natl Acad Sci USA* **2016,** *113* (13), 3533-3538.
51. Stallinga, S., Strehl ratio for focusing into biaxially birefringent media. *J Opt Soc Am A* **2004,** *21* (12), 2406-2413.
52. Turrell, G., Analysis of Polarization Measurements in Raman Microspectroscopy. *J Raman Spectrosc* **1984,** *15* (2), 103-108.
53. Bae, S. C.; Lee, H.; Lin, Z. Q.; Granick, S., Chemical imaging in a surface forces apparatus: Confocal Raman spectroscopy of confined poly(dimethylsiloxane). *Langmuir* **2005,** *21* (13), 5685-5688.
54. Nanjundiah, K.; Kurian, A.; Kaur, S.; Singla, S.; Dhinojwala, A., Crystallinelike Ordering of Confined Liquids at the Moving Contact Line. *Phys Rev Lett* **2019,** *122* (12), 128004.
55. Praveena, M.; Bain, C. D.; Jayaram, V.; Biswas, S. K., Total internal reflection (TIR) Raman tribometer: a new tool for in situ study of friction-induced material transfer. *Rsc Adv* **2013,** *3* (16), 5401-5411.
56. Contreras-Naranjo, J. C.; Ugaz, V. M., A nanometre-scale resolution interference-based probe of interfacial phenomena between microscopic objects and surfaces. *Nat Commun* **2013,** *4*, 2865.
57. Manica, R.; Parkinson, L.; Ralston, J.; Chan, D. Y. C., Interpreting the Dynamic Interaction between a Very Small Rising Bubble and a Hydrophilic Titania Surface. *J Phys Chem C* **2010,** *114* (4), 1942-1946.

58.	Radler, J.; Sackmann, E., On the Measurement of Weak Repulsive and Frictional Colloidal Forces by Reflection Interference Contrast Microscopy. *Langmuir* **1992,** *8* (3), 848-853.
59.	Shi, C.; Cui, X.; Xie, L.; Liu, Q. X.; Chan, D. Y. C.; Israelachvili, J. N.; Zeng, H. B., Measuring Forces and Spatiotemporal Evolution of Thin Water Films between an Air Bubble and Solid Surfaces of Different Hydrophobicity. *Acs Nano* **2015,** *9* (1), 95-104.
60.	Wiegand, G.; Neumaier, K. R.; Sackmann, E., Microinterferometry: three-dimensional reconstruction of surface microtopography for thin-film and wetting studies by reflection interference contrast microscopy (RICM). *Appl Optics* **1998,** *37* (29), 6892-6905.
61.	Derjaguin, B. V.; Abrikossova, I. I., Direct Measurements of Molecular Attraction of Solids. *J Phys Chem Solids* **1958,** *5* (1-2), 1-10.
62.	Dobbs, H. A.; Kaufman, Y.; Scott, J.; Kristiansen, K.; Schrader, A. M.; Chen, S. Y.; Duda, P.; Israelachvili, J. N., Ultra-Smooth, Chemically Functional Silica Surfaces for Surface Interaction Measurements and Optical/Interferometry-Based Techniques. *Adv Eng Mater* **2018,** *20* (2), 1700630.

**Table of Content figure**

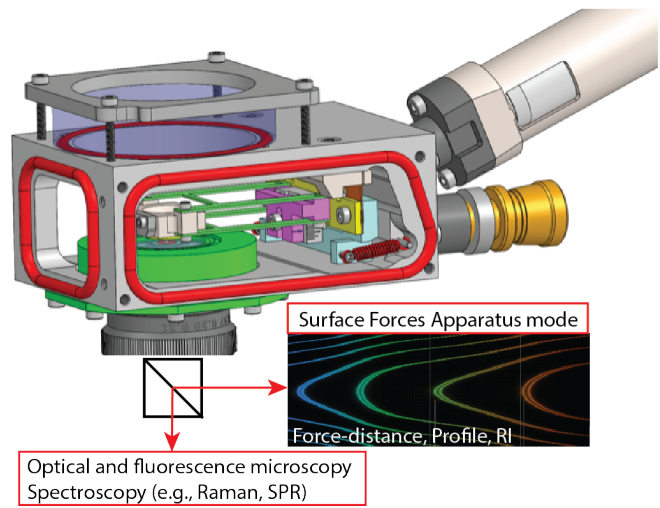